\documentstyle[amscd,amssymb,12pt]{amsart}

\def\hcorrection#1{\advance\hoffset by #1 }
\def\vcorrection#1{\advance\voffset by #1 }
\vcorrection{-.5in}
\hcorrection{-.7in}

\input xypic \xyoption{curve}
\input epsf

\begin{document}

\title{Quantum Relativity: an essay} 
\author{Lucian M. Ionescu}        % Enter your name between curly braces
\date{May 15, 2010}          % Enter your date or \today between curly braces
\address{Address: Illinois State University, Normal, IL 61790-4520;
email: LMIones@@ilstu.edu}

\begin{abstract}
Is ``Gravity'' a deformation of ``Electromagnetism''?

\vspace{.2in}
Deformation theory suggests ``quantizing'' {\em \underline{Special} Relativity}:
formulate {\em Quantum Information Dynamics} (QID) as an 
$SL_2(C)_h$-gauge theory of dynamical lattices,
with {\em unifying gauge ``group''} the quantum bundle obtained from the {\em Hopf monopole bundle}
underlying the quaternionic algebra and Dirac-Weyl spinors:
$$\xymatrix @C=1pc @R=.5pc{
S^1\cong SU_1 \rto^{fiber} & S^3\cong SU_2 \drto_{Hopf\ bundle\quad } \rrto^{2:1} & & 
SO_3\dlto^{\qquad Homogeneous\ manifold} & SO_2 \lto \\
                                 &                                 & S^2  & 
}$$
The deformation parameter is the inverse of light speed $\overline{h}=1/c$, 
in duality with Planck's constant $h$:
$$Hall\ conductivity: \quad \frac{h}{e^2}=\alpha \overline{h}\quad : ``Cosmological\ conductivity''.$$
Then mass $m$ and electric charge $q$ form a complex coupling constant $(m,q)$,
for which the quantum determinant of the quantum group $SL_2(C)_h$ 
expresses the interaction strength as a linking number 2-form:
$$Qdet
\left(
\begin{array}{cc}
q & im'\\
-im& q' 
\end{array} 
\right)=qq'-e^{-1/\alpha}\ m m'.$$ 
There is room for both Coulomb constant $k_C$ and Newton's gravitational constant $G_N$, 
exponentially weaker then the reciprocal of the fine structure constant $\alpha$:
$$\frac{G_Nm_e^2}{k_Ce^2}\approx 10^{-54}\quad \leftrightarrow \quad 
e^{-1/\alpha}\approx 10^{-59}.$$
Thus ``Gravity'' emerges already ``quantum'', in the discrete framework
of QID, based on the quantized complex harmonic oscillator: the quantized qubit.

\vspace{.2in}
All looks promising, but will the details backup this ``grand design scheme''?
\end{abstract}

\maketitle

%\clearpage

\tableofcontents
%\clearpage

% ************************************************************************
%		Introduction	500 words
% ************************************************************************
\section{Introduction}       % Enter section title between curly braces
From a Computer Science point of view 
a new paradigm emerges in Mathematical-Physics: {\em Quantum Information Dynamics} (QID) \cite{I-MPCS, QC}.
It is a physics interface naturally crystallizing from the general
principles of the {\em Digital World Theory} \cite{I-DWT},
mathematically implemented using graph complexes and their cohomology \cite{Kon,CK,I-FL,IF},
unifying classical and quantum physics.

QID emphasizes two main points.
One is that classical and quantum information are in an ``external-internal space'' duality (IE-duality), 
one information emerging from the other: bit from qubit through measurement,
and back to qubit through superpositions (e.g. quantum erasure).

The other important aspect is that the role of Electromagnetism (EM) was underestimated, even 
after the remarkable discovery of Aharonov-Bohm a. a. of the close connection
between quantum phase and {\em classical} EM as an $SU_1$-gauge theory.

A closer inspection of {\em Special Relativity} reveals that not only time is not ``absolute''
as a conclusion of a Einstein's critical analysis of the concept of synchronization,
but also the concept of ``direction'' (parallelism) is subject to the same criticism.
Therefore an $SO_3$-connection is mandatory to make sense of ``direction correlation'';
the assumption of a metric with its induced Levi-Civita connection is clearly
an extra assumption, no longer appropriate in a modern physics dominated
by quantum theory.

In conclusion \cite{I-MPCS}, EM as a $SU_1$-gauge theory is only a ``Hopf fiber'' of 
QID as a $SU_2$-Yang-Mills theory, on whatever ``space'' may be.

This yields a nice {\em Background Space-Time Independent Theory} (BIT),
from which Quantum Mechanics (QM) and Quantum Field Theory (QFT) emerge,
through an averaging over possible space-time coordinate systems, 
which are embeddings in an ambient space-time manifold ``a la'' {\em String Theory}.
and Heisenberg canonical commutation relations implementing a categorical,

\vspace{.2in}
But then, why is there ``Gravity'' to ``spoil'' this nice picture?

At the ``antipode'' of the possibility that gravitation is a global,
entropy related phenomenon \cite{Verlinde},
we will explore the idea that
$${\Large Gravity\ is\ a\ deformation\ of\ Electromagnetism.}$$

Indeed, Lorentz group is a one-parameter ``infinitesimal'' deformation of Galilei group.
In view of General Relativity, it represents the local symmetries of physical processes
(classical or quantum).
Therefore we believe that one needs to fully \underline{deform} Galilei group
into a quantum group,
{\em not} to ``quantize'' General Relativity!
As hinted above, we claim that ``quantization'' comes for free, 
from the discretization process and categorical approach.

% ***************************************************************************
%	1000 w. left		Linking Numbers		344 words
% ***************************************************************************
\section{Linking Numbers as an Interaction Model}
In the context of a discrete model, such as QID, particles are not newtonian ``points'',
but exhibit symmetries which determine how they interact.

For example the electromagnetic interaction can be expressed using linking numbers
and topological (cohomological) degree, weather the theory uses manifolds 
\cite{EMTC, NLEM, Barrett, Post, Kiehn}, 
or algebraic chains/cochains and homological algebra methods \cite{Sternberg,I-FL}.

There is a long history regarding the debate between 
describing interactions in physics in terms of particles {\em separated} from fields
(``local physics''),
for instance using Maxwell's equations in the spirit of Newton's laws
\footnote{Here $A$ is the vector potential and $\Phi$ is the electric potential.}:
$$\frac{d}{dt}(mv+qA)=F_{Lorentz}=q[E+v\times B]=-q \nabla(\Phi-qA),$$
or adopting a relational approach (``categorical physics'') 
as in Neumann's approach \cite{FromAtoE}, p.400, 
using a mutual potential $P_{ab}$ and a (geometric) inductance $M_{ab}$,
to express the magnetic force $F_{ab}$ 
or torque $C_{ab}$ between two closed electric circuits $a$ and $b$
(\cite{Germain}, p.2,5):
$$F_{ab}=-\nabla P_{ab}, \quad P_{ab}=I_a I_b\ M_{ab}, \quad  
M_{ab}=\frac{\mu}{4\pi} \int_a \int_b \frac{dl_a\cdot dl_b}{r},$$
$$C_{ab}=-I_aI_b\ L_{ab}, \quad L_{ab}=\int_a \int_b \frac{dl_a\times dl_b}{r}.$$
The double integrals are directly related to linking number \cite{EMTC}, p.114:
$$Link(a,b)=\frac1{4\pi}\int_a\int_b \frac{dr'\times(r-r')\cdot dr}{|r-r'|^3},$$
which is obtained by rewriting Biot-Savart law by converting current elements
into moving charge elements:
$$Idl=\frac{dq}{dt} dl=dq \frac{dl}{dt}=v dq.$$
Maybe the loops could also be internal $S^1$ fibers, allowing to represent 
the electric force too as a linking number, with the appropriate 
coupling constant $1/\epsilon$.

\vspace{.2in}
So, never mind Gravity, what is Electric Force!?

\vspace{.2in}
In view of the {\em duality between mass and electric charge} at the level of the 
canonical momentum $P=mv+qA$, which is the one quantized and conserved in
experiments like Aharonov-Bohm effect, linking EM and QM like Aharonov-Bohm \cite{OP}, p.384 
(see also \cite{London,VP-articles,VP-thoughts}):
$$Fluxoid: \quad\int_c (mv+qA)ds=h,$$ 
we should ``grade'' the interacting particles with their own coupling constants
corresponding to their symmetry groups,
with hindsight from particle physics:
protons $e^+$, electrons $e^-$ and hydrogen atoms $e^0$,
without the hasty assumption that $e^+=-e^-$.

% *******************************************************************
%	650 left		Newton-Coulomb Force		569 words
% *******************************************************************
\section{A Toy Model: the Newton-Coulomb Force}
Regarding Newton's law modeling the interaction of two ``pointwise'' bodies:
$$d/dt\ (mv)=F_{interactions},$$
the left hand side is additive with respect to the particles which constitute
the subsystems of the two interacting systems $S_1$ and $S_2$, 
while the right hand side
is multiplicative, corresponding to the ``edges'' of the bipartite graph
representing the two {\em separate} systems:
\begin{equation}\label{E:N}
d/dt \sum_i m_iv_i=-\nabla \sum_{i,j} \frac{q_iq_j}{r}.
\end{equation}
Here we have assumed that distance has the role of diminishing the interaction strength,
conform to a Coulomb potential law.
We also assume that there is {\em no superposition} between interacting pairs,
so that a tensorial coupling constant ``$k_{ij}$'' is a product $a_ib_j$,
which can be absorbed in the definition of the charges $q_i$. 

With only two types of {\em real} charges $+$ and $-$,
let us relax the assumption $e^-=-e^+$ as follows:
$$e^+=e-\delta/2>0, \quad e^-=-e-\delta/2<0, \qquad e>>\delta>0.$$
Indeed, with hindsight from {\em charge-parity violation}, 
if we have {\em chirality}, why not charge-conjugation violation?
And since there is a ``gravitational attraction'', 
two electrons effectively repel each other slightly  stronger then two protons
(notice the negative sign in (\ref{E:N})):
$$Electron-electron: (e^-)^2=(-e-\delta/2)^2= e^2+e\delta+\delta^2/4,$$
$$Proton-proton (e^+)^2=(e-\delta/2)^2=e^2-e\delta+\delta^2/4,$$
while the electron-proton force is attractive (built in signs!):
$$Electron-proton: e^-e^+=(-e-\delta/2)(e-\delta/2)=-(e^2-\delta^2/4).$$
This slightly shifts the origin of the real axis by $\delta$, 
breaking the ``left-right'' symmetry.

Now the resulting interaction force is consistent with a separation between a ``bulk'' electric force,
and a ``residual'', possibly gravitational-like force:
$$F=F_E+F_G, \quad F_E\approx \pm e^2.$$
What about ``neutral'' bodies?
The residual interaction between two ``neutral'' systems, 
consisting from a pair of opposite charges each, is:
$$HH-atoms\ interaction: \ (e^+)^2+(e^-)^2+2e^+e^-=\delta^2.$$
Unfortunately it is a repelling force!
Then let's treat $\delta$ as a result of renormalization,
which according to the author {\em is} a Hopf algebra deformation \cite{I-HREN},
and apply the usual Feynman trick of ``nudging'' $\delta$ into the complex plane
to avoid ``poles'':
$$Feynman\ trick\ /\ Wick\ rotation:\quad \delta \mapsto i\delta.$$
In fact ``charges'' are {\em sources} of singularities,
needed for monodromy if translating differential equations in the language 
of differential Galois theory; so ``charges'' would translate into
complex residues or periods in a theory formulating interactions via 
linking numbers and topological degrees.

Now assume that for one system with $n_+$ ``positive'' charged particles
and $n_-$ ``negative'' charged particles,
the total ``mass'' and ``net charges'' are defined as:
$$m=n_++n_-, \quad q=n_+-n_-.$$
{\em If} we take the real part of everything, with the above complex 
charge-parity violation:
$$e^+=e-i\delta/2, \quad e^-=-e-i\delta/2,$$
we recover in the Coulomb law both the electric force and the gravitational force terms:
\begin{equation}\label{E:CNF}
F_{EG}=-[e^2\ qq'-\delta^2\ mm']/r^2,
\end{equation}
with $\delta^2$ playing the role of the gravitational constant $G_N$.

% *************************************************************************************
%	100 left	Deforming Special relativity		406 words 
% *************************************************************************************
\section{Deforming Special Relativity: $SL_2(C)_h$}
\footnote{Due to limitations of the essay we will be brief here.}
One direct way to break the symmetry between positive and negative charges,
is to deform the symmetry group, so that by Noether Theorem
in the Lagrangian formalism, to get an asymmetry between 
the corresponding charges and conserved currents associated with the symmetries.

Quantum computing corresponds to Special Relativity via 
Klein correspondence of Twistor Theory \cite{Twistor-Theory} (Hermitian model):``$2+2^*=1+3$''.
The corresponding group controlling QC, SR and other models of Quantum Gravity is $SL_2(C)$,
with Lie algebra $sl_2(C)=su_2(R)+su_2$.

On the ``$2+2^*$'' side, the quaternions are a generalized complex structure \cite{GCS}:
$$H=(T^*C,\omega)=C\oplus C^*, \quad J:H\to H, J^2=-1,$$
``hosting'' qubits / $SU_2$ and the Hopf bundle (complex harmonic oscillator),
while the ``$1+3$'' side
can be viewed at the infinitesimal level as a central extension of the 
angular velocity Lie algebra \cite{Arnold}, with the cross-product as a Lie bracket:
$$R\to H\to g=(R^3,\times).$$
We interpret it as an infinitesimal deformation, with deformation parameter 
$\overline{h}=1/c$:
$$H\ni q=t+v \overline{h},\ v\in R^3.$$
Now quantize the symplectic group $H=T^*C$, as the complex version 
of the Heisenberg group viewed as a central extension:
$$R\to Heisenberg\to (T^*R,\omega),$$
by using the Backer-Campbell-Hausdorff formula (with $[,]$ in place of $\times$):
$$q\oplus q'=q+q'+\frac12[q,q']\overline{h}+\frac1{12}([q,[q,q']]-[q',[q',q]])\overline{h}^2+\dots .$$
On the ``$2+2^*$'' side it should correspond to quantizing $SL_2(C)$
as a quantum group.

%\vspace{.1in}
Surprisingly enough, if we factor Coulomb's constant in (\ref{E:CNF}), 
the coefficient in the Coulomb-Newton force
looks like the corresponding {\em quantum determinant}  \cite{HA2QG}:
$$qq'-\delta^2/e^2\ mm' \quad \leftrightarrow 
Qdet
\left(
\begin{array}{cc}
q & im'\\
-im & q' 
\end{array} 
\right)=qq'-e^{-h} \ mm'.
$$
The ``flip'' in the second column 
could be due to a complex rotation in a Kahler form.

Moreover the deformation factor $e^{-h}$ could play the role of the
universal gravitational constant, explaining why gravity
is so weak compared to EM! (see Abstract).

\vspace{.2in}
In \cite{I-QR} we suggest the possibility that the two deformation parameters,
the conductivity of interactions $\overline{h}=1/c$ and Planck's constant $h$,
are in duality:
$$Hall\ conductivity: \quad \frac{h}{e^2}=\alpha \overline{h}\quad : Cosmological\ ``conductivity''.$$
This would just express the ``amazing duality'' between 
{\em Kepler's Problem} and {\em Harmonic Oscillator} \cite{Arnold},
which we should take as a ``sign'' that micro-cosmos and macro-cosmos {\em are} indeed dual.
But is this a ``T-duality'' $\overline{h}=1/c$ or ``S-duality'' $\overline{h}=1/\alpha$,
in the spirit of String Theory \cite{ST-web-introd})?

\section{Conclusions}
The weakness of gravity is taken as a hint that gravity
is a deformation of ``Electromagnetism'', 
corresponding to a charge-conjugation violation due to non-commutativity
and chirality of the unifying gauge ``group'', the Hopf monopole bundle.
Quantizing it as a complex harmonic oscillator, allows to represent
the Newton-Coulomb force as an $SL_2(C)_h$ quantum determinant!

This is a research avenue, the author believes, it is worth pursuing.

% ***********************************************************************
% 	1846 words with formulas	Bibliography
% ***********************************************************************

% Set the ending of a LaTeX document

\begin{thebibliography}{33}

\bibitem{Arnold}  Arnold, V.I., Kozlov, V.V., Neishtadt, A.I., Iacob, A.,
Mathematical Aspects of Classical and Celestial Mechanics,
Encyclopaedia of Mathematical Sciences, 3rd ed., 2006.

\bibitem{Sternberg} Bamberg, P. and Sternberg, S.,
A course in mathematics for the students of physics Vol.2, 
Cambridge University Press, 1990.

\bibitem{Barrett} Barrett, T., Topological foundations of EM,
World Scientific Series in Contemporary Chemical Physics, Vol. 26, 2008.

\bibitem{SR21} Cacciatori, S., Gorini, V., Kamensnshchik, A.,
Special Relativity in the 21st century, arXiv:0807.3009v1, 2008.

\bibitem{QG-PC} Chari, V. and Prissley, A., A guide to quantum groups,
Cambridge University Press, 1994.

\bibitem{CK} Connes, A., Kreimer, D., 
Renormalization in quantum field theory and the Riemann-Hilbert problem, 
hep-th/9912092.

\bibitem{FromAtoE} Darrigol, O., Electrodynamics From Ampere to Einstein,
Oxford University Press, 2002.

\bibitem{IF} Fiorenza, D., Ionescu, L., 
Graph complexes in deformation quantization, LMP, Vol73, No.3, Sept. 2005, pp.193-208.

\bibitem{EMTC} Gross, P. W., and Kotiuga, P. R., Electromagnetic Theory and Computations:
a topological approach, Mathematical Sciences Research Institute Publications,
Vol. 48, 2004.

\bibitem{GCS} Gualtieri, M., Generalized Complex Geometry, arXiv: math.DG/0703298.

\bibitem{Twistor-Theory} Huggett, S. A., Tod, K. P., An introduction to twistor theory,
London Mathematical Society Student Texts, Vol.4,
Cambridge University Press, 2nd ed., 2001.

\bibitem{I-DWT} Ionescu, L. M.: 1) The Digital World Theory v.1, Olimp Press, 2005; 
2) Q++ and a Non-standard Model, 2008, lulu.com.

\bibitem{I-FL} Ionescu, L.M.: 1) The Feynman Legacy, 
Int. J. Pure and Appl. Math., Vol.48, No.3, 2008, pp.333-355, math/0701069; 
2) Cohomology of Feynman graphs and perturbative Quantum Field Theory,
Focus on Quantum Field Theory, Nova Publishers Inc., O. Kovras (editor), 
ISBN: 1-59454-126-4, 2004, 17 pages; math.QA/0506142.

\bibitem{I-MPCS} Ionescu, L.M.,
On some points of mathematical physics from a computer science
perspective, in preparation.

\bibitem{I-QR} Ionescu, L.M., Quantum Relativity, in preparation.

\bibitem{I-HREN} Ionescu, L.M., M. Marsalli, 
A Hopf algebra deformation approach to renormalization,
Applied Sciences, Vol. 10 (2008), pp.107-114; hep-th/0307112.

\bibitem{Kiehn} Kiehn, R., www22.pair.com/csdc/ 
1) Topology and topological evolution of EM fields and currents;
2) periods on manifolds, quantization and gauge; 
3) Topological torsion, chirality and the Hopf map;
4) Chirality and Helicity vs. Spin and Torsion.

\bibitem{VP-articles} Konopinski, E. J., 
What the electromagnetic vector potential describes,

\bibitem{Kon} Kontsevich, M., Deformation of Poisson manifolds I,
q-alg/9709040.

\bibitem{London} London, F., Superfluids, Vol. I:
Macroscopic Theory of Superconductivity, Dover publications,
2nd ed., 1960.

\bibitem{OP} Olariu, S., and Popescu, I. I., 
The quantum effects of electromagnetic fluxes,
Rev. Modern Pys., Vol. 57, No.2, April 1985, 339-436.

\bibitem{Post-QR} Post, E. J., Quantum reprogramming,
Kluwer Academic Publishers, 1995.

\bibitem{Post} Post, E.J.:
1) Can microphysical structure be probed by period integrals?,
Phys. Rev. D, Vol. 25, No.12, June 1982, pp. 3223-3229;
2) The constitutive map and some of its ramifications,
Annals of Physics, {\bf 71}, pp. 497-518 (1972); 
3) On the quantization of the Hall impedance, 
Phys. Lett., Vol. 94A, No.8, March 1983, pp. 343-345.

\bibitem{NLEM} Ranada, A. F. and Trueba, J. L., 
Topological EM with hidden non-linearity, Modern Nonlinear Optics, Part 3,
2nd ed.: Advances in Chemical Physics, Vol. 119, 2001.

\bibitem{Germain} Rousseaux, G., On the interaction between a current density
and a vector potential, physics/0503108.

\bibitem{VP-thoughts} Semon, M. D., Taylor, J. R.,
Thoughts on the magnetic vector potential, Am. J. Phys. {\bf 64} (11), November 1996,
1361-1369.

\bibitem{Verlinde} Verlinde, E.P.,
On the Origin of Gravity and the Laws of Newton, 1001.0785.

\bibitem{HA2QG} Timmermann, T.,
Invitation to quantum groups and duality : from Hopf algebras to multiplicative unitaries and beyond,
EMS Books in Mathematics, 2008.

\bibitem{Witten} Witten, E., From superconductors and four-manifolds to weak interactions,
Bull. AMS, Vol. 44, No.3, July 2007, pp. 361–391.

\bibitem{QC} Yanofsky, N., S., and Mannucci, M. A., Quantum Computing for Computer Scientists,
Cambridge University Press, 2008; 0708.0261.

\bibitem{ST-web-introd} Schwarz, P., So what is string theory, then?,
www.superstringtheory.com/basics.

\end{thebibliography}
\end{document}